# Theoretical Investigation of Creeping Viscoelastic Flow Transition Around a Rotating Curved Pipe

S. E. E. Hamza and Mostafa Y. El-Bakry[1]

**Abstract**— The study of creeping motion of viscoelastic fluid around a rotating rigid torus is investigated. The analysis of the problem is performed using a second-order viscoelastic model. The study is carried out in terms of the bipolar toroidal system of coordinates $(\eta, \theta, \varphi)$ where the toroid is rotating about its axis of symmetry ($z$-axis) with angular velocity $\Omega$. The problem is solved within the frame of slow flow approximation. Therefore, all variables in the governing equations are expanded in a power series of $\Omega$. A set of successive partial differential equations is obtained. The equations of motion governing the first and second-order are formulated and solved for the first-order only in this paper. However, the solution of the second-order equations will be the subject of a part two of this series of papers. Analytically, Laplace's equation is solved via the usual method of separation of variables. This method shows that, the solution is given in a form of infinite sums over Legendre functions of the first and second kinds. From the obtained solution it is found that, the leading term, $W^{(1)}(\eta,\theta)$, of the velocity which lies in the direction of the $\varphi$-coordinate represents the Newtonian flow. The second-order term shows that, the only non vanishing term is the stream function $\Psi^{(2)}(\eta,\theta)$, which describes a secondary flow in $\eta\theta$ -plane. The distribution of the surface traction at the toroid surface is calculated and discussed. Considering hydrodynamically conditions, the effects of toroidal geometrical parameters on the flow field are investigated in detail.

## 1. INTRODUCTION

The flow around a circular pipe is of importance in many engineering problems such as heat exchangers, lubrication systems, aerospace industries, chemical reactors, solar energy engineering and in bio-mechanics such as blood flow in catheterized artery, arteriography of the blood vessel, etc. Steady flow in a curved pipe in terms of small values of parameters that characters the flow was first studied by Dean [1], [2]. In later works on curved pipes, a variety of Dean numbers have been used by different researchers [3]. Accurate calculations of the steady and laminar flow in a coiled pipe of circular cross section have been carried out by Collins and Dennis [4], and Dennis [5]. They used finite difference method to solve the flow equations in intermediate range of Dean number. Also, the effect of catheterization on the flow characteristics in a curved artery was studied by Karahalios [6], Jayaraman and Tiwari [7] and Dash et al. [8]. Approximate theories at high Dean number have been presented by Barua [9] and Ito [10]. The relationship between asymptotic theory and the full results has been surveyed by Dennis and Ri-

[1] Salah Eed Ebrahim Hamza, Physics Department, Faculty of Science, Benha University, Egypt, PH-00201122688273, E-mail: salah.hamza@fsc.bu.edu.eg

Mostafa Y. El-Bakry, Physics Department, Faculty of Science, Benha University, Egypt, PH-00201016781977, E-mail: oelbakry13@yahoo.com.

ley [11] and the technical difficulties associated with the former have been outlined.

The physics of the fluid flow around a curved pipe are very complicated due to the presence of curvature generating centrifugal and pressure forces in the curvature direction. Typically the geometry that has received that most attention is that of small curvature and the steady solution structure in the limits of small and large Den number has been clarified [12]. Multiple solution branches are known to exist [13] and recent work has clarified the effects of finite pipe curvature on the steady solution branches [14] and also identified multiple periodic solutions [15].

In [16], Soh and Berger solved the Navier-Stokes equation for the fully developed flow of a homogenous Newtonian fluid in a curved pipe of circular cross-section for arbitrary curvature ratio. They solved numerically the Navier-Stokes system in the primitive variables form using a finite difference scheme. Closed form perturbation solutions for a second-order model were obtained by several authors in the special case where the second normal stress coefficient is zero. For this model, Jitchote and Robertson [17] obtained analytical solutions to the perturbation equations and analyze the effects of non-zero second normal stress coefficient on the behavior of the solution. Theoretical results regarding this problem using a splitting method were obtained by Coscia and Robertson [18].

Recently, Non-Newtonian fluid simulations in curved pipe have received some attention. Both analytical [19], [20], [21] and numerical [22], [23] approaches have been employed in these studies. Viscoelastic Dean flow in curved ducts is influenced by many factors such as centrifugal, inertia and viscous forces, curvature angle, aspect ratio and rheological properties (such as normal stress differences and nonlinear viscometric behavior). The first normal stress difference of viscoelastic fluids creates a strong hoop stress near the outer side of the curvature which strengthens the vortices while the negative second normal stress difference weakens the secondary flows intensity. Further research including experimental aspects [22], [24] has identified that the intensity of this flow is strongly linked to the toroid curvature radius and first normal stress difference. An increase in first normal stress difference induce an increase in secondary flow intensity and decreases the flow rate. The second normal stress difference has the opposite effect on the secondary flow i.e. it causes a reduction in the rate of the secondary flow, as demonstrated experimentally by Tsang and James [25] and numerically by Yanase et al. [26] in which they implemented the spectral method for flow through a slightly curved circular tube.

In the current paper, the viscoelastic creeping (very slow) flow around a rotating curved circular pipe is investigated analytically, using the second-order viscoelastic model. This rheological model is suitable for describing creeping flows of viscoelastic fluids. The governing equations can be written as two coupled, nonlinear partial differential equations for the stream function and rotational component. The format of this paper is as follows. The next section introduces the bipolar-toroidal coordinate system which is used in this paper. The formulation of the system of equations that governing the flow fields are given in section three. The method of solution is described in section four. In section five, the solution of the first order approximation is discussed. The discussion of the final results included in section six. Finally, our conclusions are given in section seven.

## 2. TOROIDAL COORDINATES

Since we are interested in studying the behavior of steady flow around a curved pipe with circular cross section, it is more convenient to use the bipolar toroidal coordinate system, in the variables ($\eta, \theta, \varphi$). The coordinates are related to the rectangular system $x, y, z$ through:

$x = h \sinh \eta \cos \varphi, \quad y = h \sinh \eta \sin \varphi, \quad z = h \sin \theta,$ (1a)

or
$$\rho = \sqrt{x^2 + y^2} = h \sinh\eta, \quad z = h \sin\theta \tag{1b}$$
where,
$$h = \frac{a}{\cosh\eta - \cos\theta}, \tag{1c}$$
is the scale factor and "$a$" is the focal distance. Toroidal coordinates are generated by rotating the circles of Fig. 1 about the z-axis, [27], and [28]. The surfaces of $\eta = const.$ are toroids, the surfaces of $\theta = const.$ are spherical bowls while the surfaces $\varphi = const.$ are half planes through the axis of symmetry, Fig. 2. The coordinates varies between $0 \leq \eta < \infty$, $0 < \theta \leq 2\pi$, $0 < \varphi \leq 2\pi$. The observations
$$(\rho - a \coth\eta)^2 + z^2 = \left(\frac{a}{\sinh\eta}\right)^2, \tag{2a}$$
$$\rho^2 + (z - a \cot\theta)^2 = \left(\frac{a}{\sin\theta}\right)^2, \tag{2b}$$
enable the orthogonal circles of constant $\eta$ and constant $\theta$ to be drown in the ($\rho$, $z$) plane as in Fig. 1. Note that, the origin of the ($\rho$, $z$) plane is mapped to the point $(\eta,\theta) = (0,\pi)$ and the point at infinity in the ($\rho$, $z$) plane is mapped to the point $(\eta,\theta) = (0,0)$ as $\eta \to \infty$, $\theta$ is ignored and $(\rho, z) = (a, 0)$.

In toroidal coordinates, the azimuthal angle denoting a rotation about the z-axis, and the distance from this axis is $\rho = h \sinh\eta$. The coordinates are chosen so that, $\eta = \eta_s$ represents the toroidal surface. Therefore, the original dimensions of the toroid $\eta_s$ are:
$$R = a \coth\eta_s, \quad \text{and} \quad r = \frac{a}{\sinh\eta_s}, \tag{3a}$$
so, the toroidal parameters, $a$ and $\eta_s$, are related to the geometric parameters $R$, $r$ of the torus as follow:
$$a = \sqrt{R^2 - r^2}, \quad \text{and} \quad \eta_s = \cosh^{-1}\frac{R}{r}, \tag{3b}$$

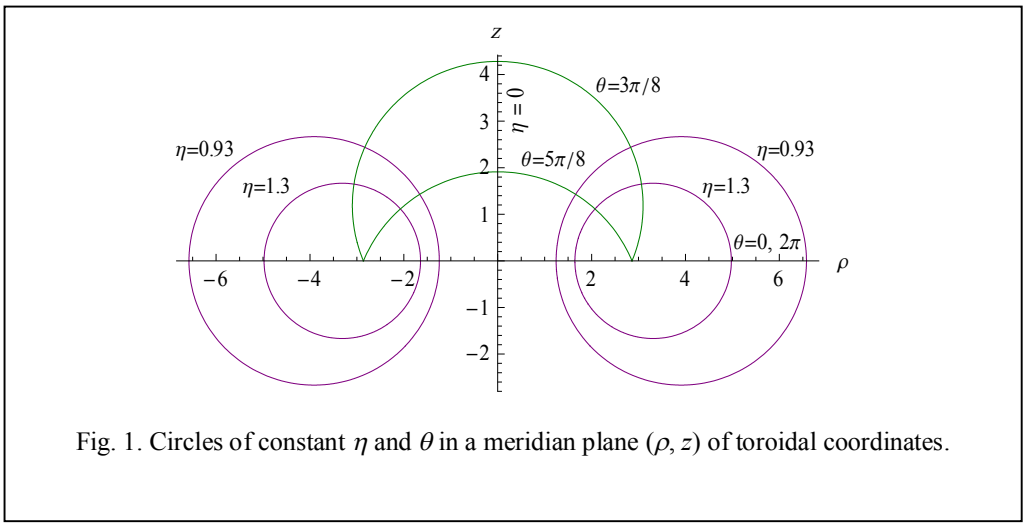

Fig. 1. Circles of constant $\eta$ and $\theta$ in a meridian plane ($\rho$, $z$) of toroidal coordinates.

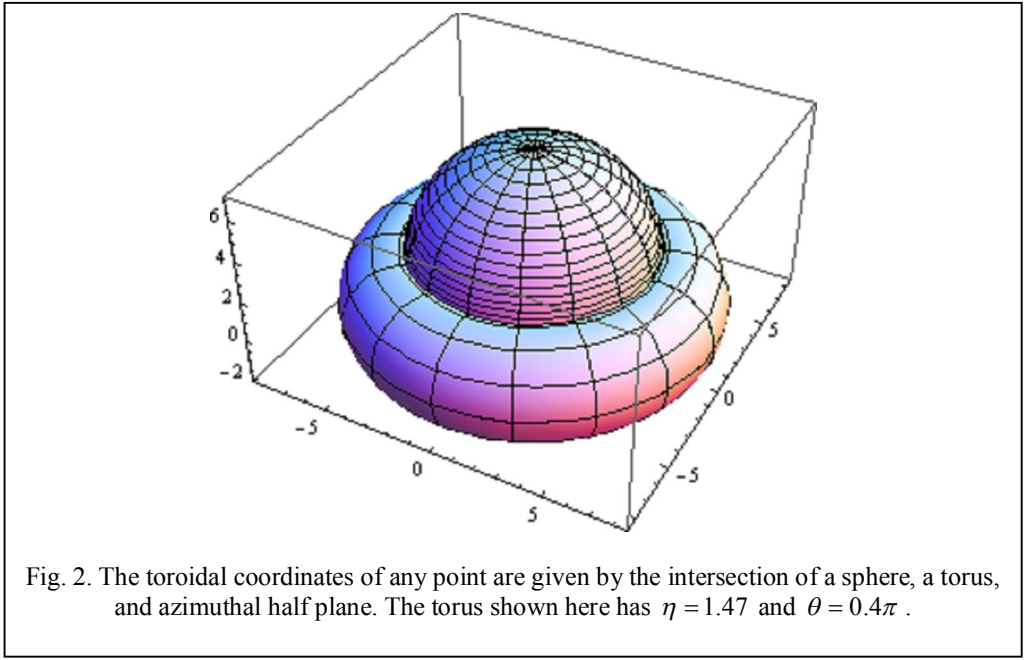

Fig. 2. The toroidal coordinates of any point are given by the intersection of a sphere, a torus, and azimuthal half plane. The torus shown here has $\eta = 1.47$ and $\theta = 0.4\pi$.

### 3. THE GOVERNING EQUATIONS

We consider a solid torus with a centerline (major) radius R and a cross-sectional (minor) radius r rotates in an infinite incompressible steady second-order viscoelastic fluid with constant angular velocity $W_s$ about its polar axis. As a result of a torus rotation, a steady axially symmetric creeping flow has been established around the torus. Figure 3 shows the rotating curved pipe and its toroidal parameters. The flow is governed by the Navier Stokes and continuity equations:

$$\nabla \cdot \underline{\dot{x}} = 0, \tag{4a}$$

$$\rho(\underline{\dot{x}} \cdot \nabla \underline{\dot{x}}) = -\nabla p + \nabla \cdot \underline{\underline{\tau}}, \tag{4b}$$

here $\underline{\dot{x}}$ is the velocity vector, $p$ is the pressure and $\underline{\underline{\tau}}$ is the stress tensor. Assume that the forces due to viscoelasticity are dominate such that the inertial term, $\underline{\dot{x}} \cdot \nabla \underline{\dot{x}}$, is negligible order. Thus,

$$-\nabla p + \nabla \cdot \underline{\underline{\tau}} = 0. \tag{4c}$$

The constitutive equation for a second-order fluid, as suggested by Coleman and Noll [29] relates the stress tensor $\underline{\underline{\tau}}$ to the kinematical Rivilin-Ericksen tensor $\underline{\underline{A}}_k$ by the expression

$$\underline{\underline{\tau}} = \mu \underline{\underline{A}}_1 + \alpha_1 \underline{\underline{A}}_2 + \alpha_2 \underline{\underline{A}}_1^2, \tag{5a}$$

where $\mu$ is the coefficient of viscosity and $\alpha_1$ and $\alpha_2$ are two second-order material coefficients related to the normal stress differences. The tensors $\underline{\underline{A}}_1$ and $\underline{\underline{A}}_2$ stands for the first two Rivilin-Ericksen tensor defined by:

$$\underline{\underline{A}}_1 = \nabla \underline{\dot{x}} + (\nabla \underline{\dot{x}})^T, \tag{5b}$$

and

$$\underline{\underline{A}}_2 = \underline{\dot{x}} \cdot \nabla \underline{\underline{A}}_1 + \nabla \underline{\dot{x}} \cdot \underline{\underline{A}}_1 + \left(\nabla \underline{\dot{x}} \cdot \underline{\underline{A}}_1\right)^T, \qquad (5c)$$

The material constants must satisfy the following restrictions:
$$\mu \geq 0, \ \alpha_1 \geq 0, \ \alpha_1 + \alpha_2 = 0, \qquad (5d)$$

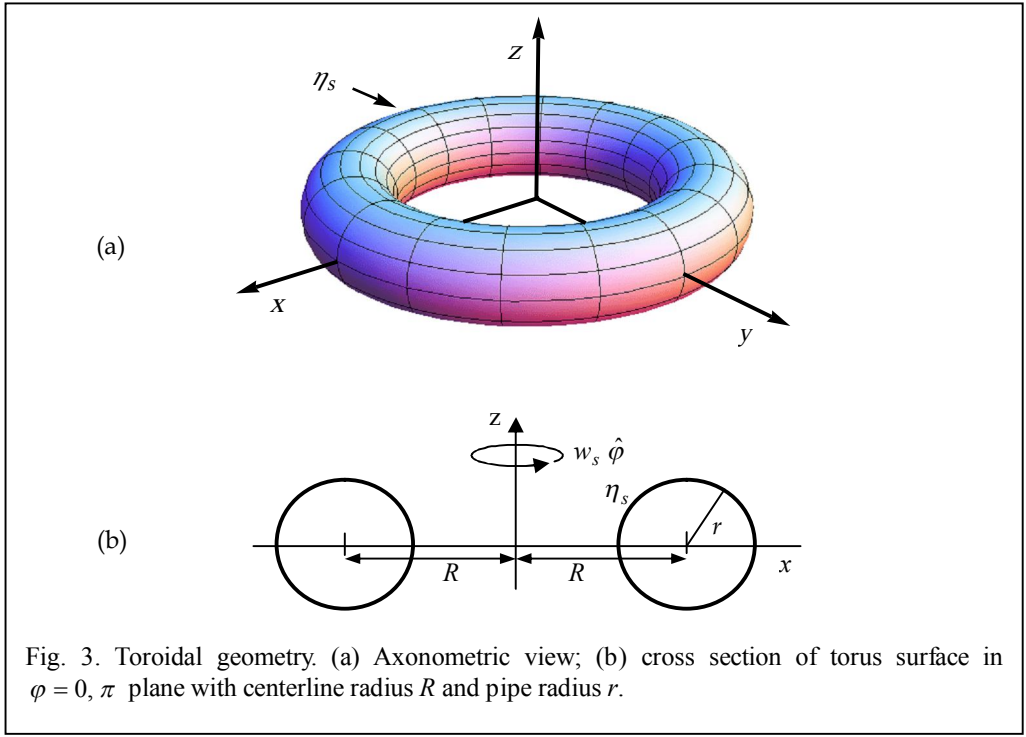

Fig. 3. Toroidal geometry. (a) Axonometric view; (b) cross section of torus surface in $\varphi = 0, \pi$ plane with centerline radius $R$ and pipe radius $r$.

Symmetry about the z-axis implies that the velocity field is independent on the coordinate $\varphi$. Thus,
$$\underline{\dot{x}} = U(\eta,\theta)\hat{\eta} + V(\eta,\theta)\hat{\theta} + W(\eta,\theta)\hat{\varphi}, \qquad (6)$$
since the flow is described in the meridian plane, the velocity components $U$ and $V$ can be expressed in terms of the stream function $\Psi(\eta,\theta)$, which satisfy the continuity equation as:
$$\underline{V}_\perp = U\hat{\eta} + V\hat{\theta} = \frac{1}{h^2 \sinh\eta}\left(\Psi_{,\theta}\hat{\eta} - \Psi_{,\eta}\hat{\theta}\right) = \nabla \wedge \left(\frac{\Psi\hat{\varphi}}{h\sinh\eta}\right), \qquad (7a)$$
so
$$\underline{\dot{x}} = \underline{V}_\perp + W\hat{\varphi} = \nabla \wedge \left(\frac{\Psi\hat{\varphi}}{h\sinh\eta}\right) + W\hat{\varphi}. \qquad (7b)$$
Taking the divergence of (5a) then substituting in (4c), we get
$$\mu\nabla^2\underline{\dot{x}} + \underline{\Lambda} - \nabla p = 0. \qquad (8a)$$
where
$$\underline{\Lambda} = \nabla \cdot \left(\alpha_1 \underline{\underline{A}}_2 + \alpha_2 \underline{\underline{A}}_1^2\right), \qquad (8b)$$
Since $\underline{\dot{x}} = \underline{V}_\perp + W\hat{\varphi}$ and $\underline{\Lambda} = \underline{\Lambda}_\perp + \Lambda_3\hat{\varphi}$ with $\underline{V}_\perp = U\hat{\eta} + V\hat{\theta}$ and $\underline{\Lambda}_\perp = \Lambda_1\hat{\eta} + \Lambda_2\hat{\theta}$, then (8a) may be decomposed into the $\varphi$- component and the vector equation including the $\eta$ - and $\theta$ -

components as:
$$\nabla^2(W\hat{\varphi}) + \Lambda_3 \hat{\varphi} = 0, \tag{9a}$$
$$\nabla^2 \underline{V}_\perp + \underline{\Lambda}_\perp - \nabla p = 0. \tag{9b}$$

Applying the curl operation to (9b) and using (7a), we get:
$$\nabla^4\left(\frac{\Psi\hat{\varphi}}{h\sinh\eta}\right) - \frac{\hat{\varphi}}{h^2}\left[\partial_\eta(h\Lambda_2) - \partial_\theta(h\Lambda_1)\right] = 0, \tag{9c}$$

We assume that, the torus is rotating about z-axis with constant velocity $W_s\hat{\varphi}$. Therefore the linear velocity at infinity ($\eta \to 0$) vanishes, $\underline{\dot{x}}(0,\theta) = 0$, while at the toroid surface is $\underline{\dot{x}}(\eta_s,\theta) = W_s\hat{\varphi}$. So the boundary conditions take the form:
$$W = \begin{Bmatrix} W_s \\ 0 \end{Bmatrix}, \quad \Psi = \Psi_{,\eta} = \begin{Bmatrix} 0 \\ 0 \end{Bmatrix} \quad \text{at} \quad \eta = \begin{Bmatrix} \eta_s \\ 0 \end{Bmatrix}. \tag{10}$$

## 4. APPROXIMATE SOLUTION FOR THE VELOCITY FIELD

The solution of the problem reduces to the determination of the scalar components $W$ and $\Psi$ such that boundary conditions (10) are satisfied. The solution is obtained by the perturbation method. This method may be summarized into the following step:

1. For small $\Omega$ the dynamical functions $W$ and $\Psi$ can be expanded into power series about $\Omega = 0$ [30] as
$$W = \sum_{k=1} \Omega^k W^{(k)} + O(\Omega^m), \tag{11a}$$
$$\Psi = \sum_{k=1} \Omega^k \Psi^{(k)} + O(\Omega^m), \tag{11b}$$

2. Substitution from (11) into (5a) gives an expression for the stress tensor into powers of $\Omega$.

3. After carrying out the decomposition, the result can be substituted into the pair of equations (9a) and (9c).

4. Equating the coefficients of equal powers of $\Omega$ produces a set of successive partial differential equations for the determination of the velocity components and the stream functions in successive order.

The previous introduction to the expansion technique allows the expansion of (5a), (9a) and (9c) in the form
$$\sum_{k=1} \Omega^k \left[\underline{\underline{\tau}}^{(k)} - \mu \underline{\underline{A}}_1^{(k)} - \alpha_1 \underline{\underline{A}}_2^{(k)} - \alpha_2 \left(\underline{\underline{A}}_1^2\right)^{(k)}\right] = 0, \tag{12a}$$
$$\sum_{k=1} \Omega^k \left[\nabla^2(W^{(n)}\hat{\varphi}) + \Lambda_3^{(k)}\hat{\varphi}\right] = 0, \tag{12b}$$
$$\sum_{k=1} \Omega^k \left\{\nabla^4\left(\frac{\Psi^{(n)}\hat{\varphi}}{h\sinh\eta}\right) - \frac{\hat{\varphi}}{h^2}\left[\partial_\eta(h\Lambda_2^{(k)}) - \partial_\theta(h\Lambda_1^{(k)})\right]\right\} = 0, \tag{12c}$$

The boundary conditions, (10), can be written as:
$$W^{(k)} = \begin{Bmatrix} W_s \delta_{1k} \\ 0 \end{Bmatrix}, \Psi^{(k)} = \Psi_{,\eta}^{(k)} = \begin{Bmatrix} 0 \\ 0 \end{Bmatrix} \quad \text{for} \quad \eta = \begin{Bmatrix} \eta_s \\ 0 \end{Bmatrix}. \tag{12d}$$

where $\delta_{ij}$ is the Kronecker delta function.

## 5. Solution of the First-Order Approximation

As usual, the solution of the first-order, $k=1$, produces the leading terms in the expansion of $W$, and $\Psi$. The solution of these terms represent the creeping flow around the rotating torus. The lowest order in (12) are:

$$\underline{\underline{\tau}}^{(1)} - \mu \underline{\underline{A}}_1^{(1)} = 0, \tag{13a}$$

$$\nabla^2 \left( W^{(1)} \hat{\varphi} \right) = 0, \tag{13b}$$

$$\nabla^4 \left( \frac{\Psi^{(1)} \hat{\varphi}}{h \sinh \eta} \right) = 0, \tag{13c}$$

with the boundary conditions:

$$W^{(1)} = \begin{Bmatrix} W_s \\ 0 \end{Bmatrix}, \quad \Psi^{(1)} = \Psi_{,\eta}^{(1)} = \begin{Bmatrix} 0 \\ 0 \end{Bmatrix}, \quad \text{for} \quad \eta = \begin{Bmatrix} \eta_s \\ 0 \end{Bmatrix}, \tag{13d}$$

The boundary conditions, (13d), imposed on (13c) implies that the only solution satisfying this boundary value problem is:

$$\Psi^{(1)} = 0. \tag{14}$$

Equation (13b) takes the form

$$\left( \nabla^2 - \frac{1}{h^2 \sinh^2 \eta} \right) W^{(1)} = 0, \tag{15a}$$

or

$$W_{,\eta\eta}^{(1)} + \frac{h}{a \sinh \eta} (1 - \cosh \eta \cos \theta) W_{,\eta}^{(1)} + W_{,\theta\theta}^{(1)} - \frac{h}{a} \sin \theta \, W_{,\theta}^{(1)} - \frac{W^{(1)}}{\sinh^2 \eta} = 0 \tag{15b}$$

this equation is R-separable in toroidal coordinate, details of the separated differential equations are outlined in several texts [27], [28]. Let

$$W^{(1)} = \sqrt{\cosh \eta - \cos \theta} \; f_1(\eta) \, f_2(\theta), \tag{16a}$$

so, the separated equations are:

$$(\xi^2 - 1) f_{1,\eta\eta} + 2\xi f_{1,\eta} - \left[ \frac{q^2}{\xi^2 - 1} + p^2 - \frac{1}{4} \right] f_1 = 0, \tag{16b}$$

$$f_{2,\theta\theta} + p^2 f_2 = 0, \tag{16c}$$

with $\xi = \cosh \eta$. The solutions of (16b) and (16c) are:

$$f_1(\xi) = \begin{pmatrix} P_{p-\frac{1}{2}}^q \\ Q_{p-\frac{1}{2}}^q \end{pmatrix} (\xi), \quad f_2(\theta) = \begin{pmatrix} \sin \\ \cos \end{pmatrix} p\theta. \tag{16d}$$

From (16d), the solutions are the associated Legendre function (toroidal function) of the first and second kinds $P_{p-\frac{1}{2}}(\cosh \eta)$ and $Q_{p-\frac{1}{2}}(\cosh \eta)$ with a parameter $p$. Therefore, the solutions are products of $\sqrt{\cosh \eta - \cos \theta} \; P_{p-\frac{1}{2}}(\cosh \eta)$ or $Q_{p-\frac{1}{2}}(\cosh \eta)$ and $\sin p\theta$ or $\cos p\theta$. In boundary value problems involving the flow around a torus, the parameter $p$ is determined by the requirement that the solution be periodic in $\theta$. Therefore, (16b) has particular solutions of the form:

$$W^{(1)} = \sqrt{\cosh\eta - \cos\theta} \sum_{p=0}^{\infty} \left[A_p \cos p\theta + B_p \sin p\theta\right] \begin{pmatrix} P_{p-\frac{1}{2}} \\ Q_{p-\frac{1}{2}} \end{pmatrix} (\cosh\eta), \tag{17a}$$

where $A_p$ and $B_p$ are arbitrary constants. In (17a), the upper row pertains to the interior problem ($\eta_s < \eta \leq \infty$) and the lower row to the exterior problem ($0 \leq \eta < \eta_s$), [31].

Due to the boundary conditions in (13d) and the boundedness requirements, only $P_{p-\frac{1}{2}}(\cosh\eta)$ and $\cos p\theta$ survive. Therefore, the solution can be written as:

$$W^{(1)} = \sqrt{\cosh\eta - \cos\theta} \sum_{p=0}^{\infty} A_p \frac{P_{p-\frac{1}{2}}(\cosh\eta)}{P_{p-\frac{1}{2}}(\cosh\eta_s)} \cos p\theta. \tag{17b}$$

We apply the first condition in (13d) to get $A_p$. One may make use of the integral

$$\int_0^{2\pi} \frac{\cos p\theta}{\sqrt{\cosh\eta - \cos\theta}} d\theta = 2\sqrt{2}\, Q_{p-\frac{1}{2}}(\cosh\eta), \tag{17c}$$

to show that

$$A_p = \frac{2\sqrt{2}}{\pi(1+\delta_{po})} \frac{Q_{p-\frac{1}{2}}(\cosh\eta_s)}{P_{p-\frac{1}{2}}(\cosh\eta_s)}. \tag{17d}$$

The solution of (15b), thus, is

$$W^{(1)} = \sqrt{\cosh\eta - \cos\theta} \sum_{p=0}^{\infty} \frac{2\sqrt{2}}{\pi(1+\delta_{po})} \frac{Q_{p-\frac{1}{2}}(\cosh\eta_s)}{P_{p-\frac{1}{2}}(\cosh\eta_s)} P_{p-\frac{1}{2}}(\cosh\eta) \cos p\theta. \tag{17e}$$

Equation (17e) represents an expression for the creeping flow around the rotating torus.

**6. SECOND-ORDER APPROXIMATION**

This step of approximation requires the determination of the kinematics tensors up to the order $O(\Omega^3)$. Therefore, the contributions of $\underline{\underline{A}}_2$ and $\underline{\underline{A}}_1^2$ have to be evaluated up to the needed order. The contributions of these tensors are expressed in terms of $W^{(1)}$ which is already calculated in the first-order approximation. This step of approximation leads to partial differential equations governing the second-order terms $W^{(2)}$ and $\Psi^{(2)}$. Taking $k = 2$ in (12), the coefficients of $\Omega^2$ are:

$$\underline{\underline{\tau}}^{(2)} - \mu \underline{\underline{A}}_1^{(2)} = 0, \tag{18a}$$

$$\nabla^2 \left(W^{(2)} \hat{\varphi}\right) + \Lambda_3^{(1)} \hat{\varphi} = 0, \tag{18b}$$

$$\nabla^4 \left(\frac{\Psi^{(2)} \hat{\varphi}}{h \sinh\eta}\right) - \frac{\hat{\varphi}}{h^2} \left[\partial_\eta \left(h\Lambda_2^{(1)}\right) - \partial_\theta \left(h\Lambda_1^{(1)}\right)\right] = 0, \tag{18c}$$

with the boundary conditions:

$$W^{(2)} = \Psi^{(1)} = \Psi_{,\eta}^{(2)} = \begin{Bmatrix} 0 \\ 0 \end{Bmatrix}, \quad \text{for} \quad \eta = \begin{Bmatrix} \eta_s \\ 0 \end{Bmatrix}, \tag{18d}$$

Equation of motion up to this order of approximation requires the determination the components $\Lambda_i$, $i = 1, 2, 3$ of the vector $\underline{\Lambda}$, in (8b), up to $O(\Omega^3)$.

## 6.1 Calculation of $\nabla \cdot \underline{\underline{A}}_2$

Let $\nabla \underline{\dot{x}} = \underline{\underline{L}}$ and $\underline{\dot{x}} = W^{(1)}\hat{\varphi}$ then

$$\underline{\underline{A}}_1 = \underline{\underline{L}} + \underline{\underline{L}}^T,$$

$$\underline{\underline{A}}_2 = \underline{\dot{x}} \cdot \nabla \underline{\underline{A}}_1 + \underline{\underline{L}} \cdot \underline{\underline{A}}_1 + (\underline{\underline{L}} \cdot \underline{\underline{A}}_1)^T,$$

$$\nabla \cdot \underline{\underline{A}}_2 = \underline{\underline{L}} : \nabla \underline{\underline{A}}_1 + (\underline{\dot{x}} \cdot \nabla)\nabla \cdot \underline{\underline{A}}_1 + \nabla^2 \underline{\dot{x}} \cdot \underline{\underline{A}}_1 + \underline{\underline{L}}^T : \nabla \underline{\underline{A}}_1 + \nabla \cdot \underline{\underline{A}}_1 \cdot \underline{\underline{L}}^T + \underline{\underline{A}}_1 : \nabla \underline{\underline{L}}^T$$

noting that

$$\nabla \cdot \underline{\dot{x}} = \nabla^2 (W^{(1)}\hat{\varphi}) = 0,$$

therefore, $\nabla \cdot \underline{\underline{A}}_2$ becomes

$$\nabla \cdot \underline{\underline{A}}_2 = \underline{\underline{L}} : \nabla \underline{\underline{A}}_1 + \underline{\underline{L}}^T : \nabla \underline{\underline{A}}_1 + \underline{\underline{A}}_1 : \nabla \underline{\underline{L}}^T$$

or in a more convenient form

$$\nabla \cdot \underline{\underline{A}}_2 = 2\underline{\underline{L}} : \nabla \underline{\underline{L}} + \frac{1}{2}\nabla(\underline{\underline{A}}_1 : \underline{\underline{A}}_1). \tag{19a}$$

## 6.2 Calculation of $\nabla \cdot \underline{\underline{A}}_1^2$

$$\nabla \cdot \underline{\underline{A}}_1^2 = \nabla \underline{\underline{A}}_1 : \underline{\underline{A}}_1 + \underline{\underline{A}}_1 : \nabla \underline{\underline{A}}_1$$

since $\nabla \cdot \underline{\dot{x}} = \nabla^2 (W^{(1)}\hat{\varphi}) = 0$, so $\nabla \underline{\underline{A}}_1 \vdots \underline{\underline{A}}_1 = 0$. Therefore, $\nabla \cdot \underline{\underline{A}}_1^2$ reduces to $\nabla \cdot \underline{\underline{A}}_1^2 = \underline{\underline{A}}_1 : \nabla \underline{\underline{A}}_1$ or

$$\nabla \cdot \underline{\underline{A}}_1^2 = 2\underline{\underline{L}} : \nabla \underline{\underline{L}} + \frac{1}{4}\nabla(\underline{\underline{A}}_1 : \underline{\underline{A}}_1). \tag{19b}$$

The substitution of (19a) and (19b) into (8b) gives

$$\underline{\Lambda}^{(1)} = \alpha_1 \nabla \cdot \underline{\underline{A}}_2 + \alpha_2 \nabla \cdot \underline{\underline{A}}_1^2$$
$$= 2(\alpha_1 + \alpha_2)\underline{\underline{L}} : \nabla \underline{\underline{L}} + \frac{1}{4}(2\alpha_1 + \alpha_2)\nabla(\underline{\underline{A}}_1 : \underline{\underline{A}}_1) + O(\Omega^3) \tag{20a}$$

The expression $\underline{\underline{L}} : \nabla \underline{\underline{L}}$ possesses the final form:

$$\underline{\underline{L}} : \nabla \underline{\underline{L}} = h^{-2} \left[ \frac{(\cosh \eta \cos \theta - 1)}{a \sinh \eta} \hat{\eta} + \frac{\sin \theta}{a} \hat{\theta} \right] F(\eta, \theta) \tag{20b}$$

with

$$F(\eta, \theta) = (W_{,\eta})^2 + (W_{,\theta})^2 + \frac{2h(\cosh \eta \cos \theta - 1)}{a \sinh \eta} W W_{,\eta} + \frac{2h \sin \theta}{a} W W_{,\theta} + \frac{1}{\sinh^2 \eta} W^2 \tag{20c}$$

The vector $\underline{\Lambda}$ in (20a) is entirely in $\eta\theta$-surface, so its components are:

$$\Lambda_1^{(1)} = \hat{\eta} \cdot \underline{\Lambda}^{(1)} = h^{-2} \left[ \frac{(\cosh \eta \cos \theta - 1)}{a \sinh \eta} \right] F(\eta, \theta), \tag{21a}$$

$$\Lambda_2^{(1)} = \hat{\theta} \cdot \underline{\Lambda}^{(1)} = h^{-2} \left[ \frac{\sin \theta}{a} \right] F(\eta, \theta), \tag{21b}$$

$$\Lambda_3^{(1)} = \hat{\varphi} \cdot \underline{\Lambda}^{(1)} = 0, \tag{21c}$$

The governing equations of the second-order approximation, (18), are decomposed into the following two boundary value problems; namely

$$\nabla^2 \left( W^{(2)} \hat{\varphi} \right) = 0, \tag{22a}$$

$$\nabla^4 \left( \frac{\Psi^{(2)} \hat{\varphi}}{h \sinh \eta} \right) = \frac{2(\alpha_1 + \alpha_2)}{\mu h^2} \left\{ \partial_\eta \left( \frac{\sin \theta}{ah} \right) F(\eta, \theta) - \partial_\theta \left( \frac{(\cosh \eta \cos \theta - 1)}{a h \sinh \eta} \right) F(\eta, \theta) \right\} \hat{\varphi} = 0 \tag{22b}$$

The boundary conditions, (18d), imposed on (22a) implies that the only solution satisfying this boundary value problem is trivial solution, $W^{(2)} = 0$, and the density function on the right hand side of (22b) which includes the first-order velocity $W^{(1)}(\eta, \theta)$ and its derivatives possesses a very complicated form. In fact, this expression is of the form of a double summation since $W^{(1)}(\eta, \theta)$ is given in the form of an infinite series. Therefore, the solution of the second-order system, (22), will be the subject of the second paper.

## 7. SURFACE TRACTION

This section is devoted to the evaluation of the surface traction at the boundary $\eta_s$. The surface traction is defined by

$$\underline{S}(\eta, \theta) \big|_{\eta_s} = \underline{\underline{\tau}}(\eta, \theta) \cdot \hat{\eta} \big|_{\eta_s} = \left( \mu \underline{\underline{A}}_1 + \alpha_1 \underline{\underline{A}}_2 + \alpha_2 \underline{\underline{A}}_1^2 \right) \cdot \hat{\eta} \big|_{\eta_s} \tag{23}$$

where the unit vector $\hat{\eta}$ is the normal to any arbitrary surface, $\eta = const$. Hence, the surface traction is the stress vector per unit area on the surface of a toroidal shell $\eta = \eta_s$.

### 7.1 Determination of the Rivlin-Ericksen Tensors

**The tensor** $\underline{\underline{A}}_1$

The velocity field $\underline{\dot{x}}$ is defined by

$$\underline{\dot{x}} = W^{(1)}(\eta, \theta) \hat{\varphi}, \tag{24}$$

therefore,

$$\nabla \underline{\dot{x}} = h^{-1} \left( \hat{\eta} \partial_\eta + \hat{\theta} \partial_\theta + \frac{\hat{\varphi}}{\sinh \eta} \partial_\varphi \right) \left( W^{(1)} \hat{\varphi} \right) \tag{25b}$$

$$= a_{13} \hat{\eta} \hat{\varphi} + a_{23} \hat{\theta} \hat{\varphi} + a_{31} \hat{\varphi} \hat{\eta} + a_{32} \hat{\varphi} \hat{\theta}$$

where

$$\left. \begin{array}{l} a_{13} = h^{-1} W^{(1)}_{,\eta} \\ a_{23} = h^{-1} W^{(1)}_{,\theta} \\ a_{31} = \dfrac{(\cosh \eta \cos \theta - 1)}{a \sinh \eta} W^{(1)} \\ a_{32} = \dfrac{\sin \theta}{a} W^{(1)} \end{array} \right\}, \tag{25b}$$

and

$$\left( \nabla \underline{\dot{x}} \right)^T = b_{13} \hat{\eta} \hat{\varphi} + b_{23} \hat{\theta} \hat{\varphi} + b_{31} \hat{\varphi} \hat{\eta} + b_{32} \hat{\varphi} \hat{\theta}, \tag{26a}$$

where

$$b_{13} = \frac{(\cosh\eta\cos\theta - 1)}{a\sinh\eta} W^{(1)}$$

$$b_{23} = \frac{\sin\theta}{a} W^{(1)}$$

$$b_{31} = h^{-1} W_{,\eta}^{(1)}$$

$$b_{32} = h^{-1} W_{,\theta}^{(1)}$$

(26b)

Therefore, Rivlin-Ericksen tensor $\underline{\underline{A}}_1$ is given by the expression:

$$\underline{\underline{A}}_1 = C_{13}(\hat{\eta}\hat{\varphi} + \hat{\varphi}\hat{\eta}) + C_{23}(\hat{\theta}\hat{\varphi} + \hat{\varphi}\hat{\theta}), \tag{27a}$$

where

$$C_{13} = C_{31} = h^{-1} W_{,\eta}^{(1)} + \frac{(\cosh\eta\cos\theta - 1)}{a\sinh\eta} W^{(1)}$$

$$C_{23} = C_{32} = h^{-1} W_{,\theta}^{(1)} + \frac{\sin\theta}{a} W^{(1)}$$

(27b)

**The tensor $\underline{\underline{A}}_2$**

The tensor $\underline{\underline{A}}_2$ possesses the form

$$\underline{\underline{A}}_2 = \underline{\dot{x}} \cdot \nabla \underline{\underline{A}}_1 + \underline{\underline{A}}_1 \cdot (\nabla \underline{\dot{x}})^T + \left[\underline{\underline{A}}_1 \cdot (\nabla \underline{\dot{x}})^T\right]^T, \tag{28}$$

we calculate each term of $\underline{\underline{A}}_2$ separately and the final form is

$$\underline{\underline{A}}_2 = Q_{11}\hat{\eta}\hat{\eta} + Q_{12}\hat{\eta}\hat{\theta} + Q_{21}\hat{\theta}\hat{\eta} + Q_{22}\hat{\theta}\hat{\theta}, \tag{29a}$$

where

$$Q_{11} = \frac{4(\cosh\eta\cos\theta - 1)}{a\sinh\eta} W^{(1)} C_{13}$$

$$Q_{12} = Q_{21} = \frac{2\sin\theta}{a} W^{(1)} C_{13} + \frac{2(\cosh\eta\cos\theta - 1)}{a\sinh\eta} W^{(1)} C_{23}$$

$$Q_{22} = \frac{4\sin\theta}{a} W^{(1)} C_{23}$$

(29b)

**The tensor $\underline{\underline{A}}_1^2$**

The tensor $\underline{\underline{A}}_1^2$ is given by using (27a) as

$$\underline{\underline{A}}_1^2 = G_{11}\hat{\eta}\hat{\eta} + G_{12}\hat{\eta}\hat{\theta} + G_{21}\hat{\theta}\hat{\eta} + G_{22}\hat{\theta}\hat{\theta} + G_{33}\hat{\varphi}\hat{\varphi}, \tag{30}$$

where

$$G_{11} = C_{13}^2$$

$$G_{12} = G_{21} = C_{13} C_{23}$$

$$G_{22} = C_{23}^2$$

$$G_{33} = C_{23}^2 + C_{13}^2$$

(30a)

So, the surface traction due to the tensors $\underline{\underline{A}}_1$, $\underline{\underline{A}}_2$ and $\underline{\underline{A}}_1^2$ is given as:

$$S(\eta,\theta) = \left(\mu \underline{\underline{A}}_1 + \alpha_1 \underline{\underline{A}}_2 + \alpha_2 \underline{\underline{A}}_1^2\right)\cdot \hat{\eta}$$
$$= S_\eta \hat{\eta} + S_\theta \hat{\theta} + S_\varphi \hat{\varphi} \tag{31a}$$

where

$$\left.\begin{array}{l} S_\eta = \alpha_1 Q_{11} + \alpha_2 G_{11} \\ S_\theta = \alpha_1 Q_{21} + \alpha_2 G_{21} \\ S_\varphi = \mu C_{13} \end{array}\right\}, \tag{31b}$$

## 8. RESULTS AND DISCUSSION

Here, we consider the problem of determining the velocity field of a second-order viscoelastic fluid due to a steady rotation of a rigid toroidal body along its symmetry axis, which is assumed to coincide with the axis $Oz$ of the cylindrical coordinates. The solution of (13b) which satisfies the boundary conditions (13d) is given in (17e). The creeping flow is one in which fluid particles are carried in toroids eccentric with the rotating torus surface $\eta_s$.

Figures 4a, 4b and 4c display the patterns of axial velocity $W^{(1)}$ contours in the vicinity of a torus for $r=1$ and $R = \frac{7}{2}$, $\frac{5}{2}$ and $\frac{3}{2}$, respectively. The contours near the torus surface, $\eta \to \eta_s$, are completely different from that in the core flow, $\eta \to 0$, (the velocity approaches zero in the core flow). It depends on $\eta$ only in the core and on $\eta$ and $\theta$ in the neighborhood of the wall. Therefore, such fluid exhibits boundary layer effects.

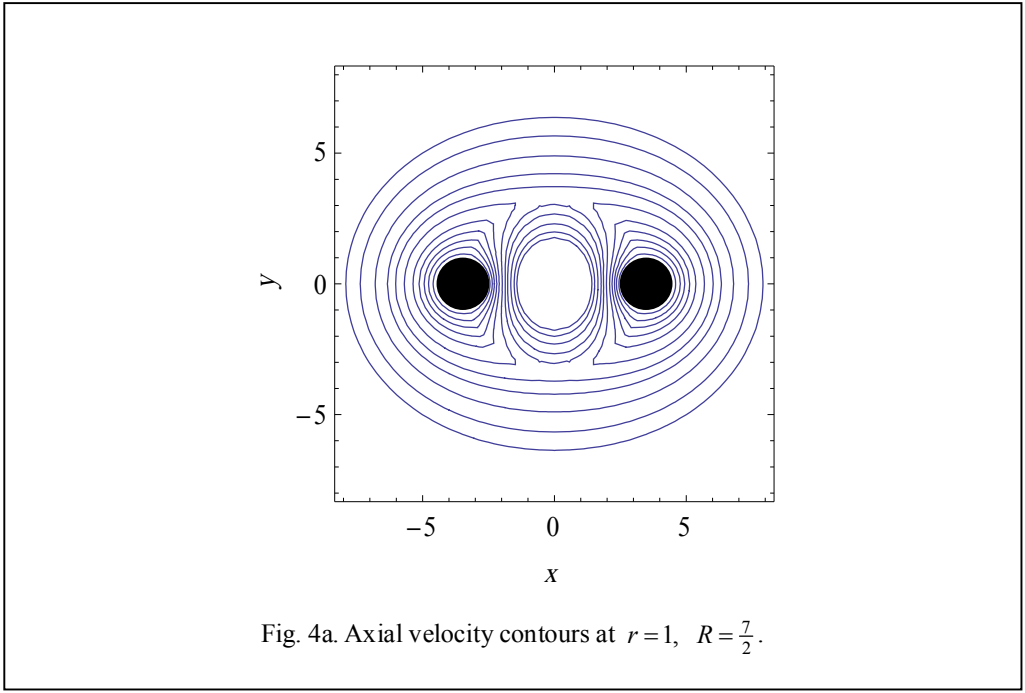

Fig. 4a. Axial velocity contours at $r=1$, $R = \frac{7}{2}$.

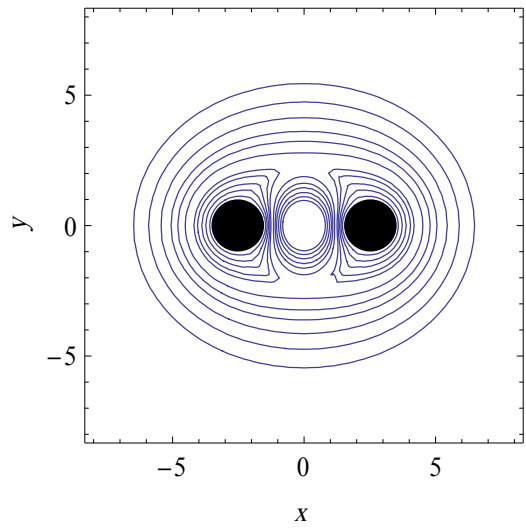

Fig. 4b. Axial velocity contours at $r=1$, $R=\frac{5}{2}$.

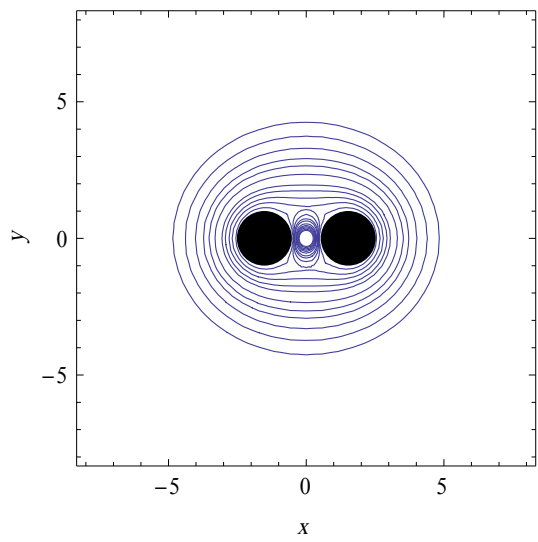

Fig. 4c. Axial velocity contours at $r=1$, $R=\frac{3}{2}$.

The behavior of $W^{(1)}$ as a function of the coordinate $\eta$ at $R=\frac{7}{2}$, $\frac{5}{2}$ and $\frac{3}{2}$ is shown in Figs. 5a, 5b and 5c respectively. It is clear that, all curves satisfy the boundary condition at the torus surface $\eta = \eta_s$. As $\eta \to 0$, the core velocities are everywhere reduce with increasing $\theta$.

Figures 6 show the behavior of $W^{(1)}$ as a function of $\theta$ for different values of $\eta$. It is clear that, $W^{(1)}$ increases with increasing $\theta$ until reaching maxima at $\theta = \pi$ then decreases as $\theta \to 2\pi$. As the value of $\eta$ decreases (core flow), the maxima decrease. The same behavior is observed for different values of toroidal geometrical parameter $R$. To give more insight about the velocity fields around the torus, the axial velocity $W^{(1)}$ is plotted in three dimensional graphs as in Fig. 7.

In experimental arrangements, the forces are measured at the surface of the torus. Therefore, the surface traction component $S_\varphi$ is calculated on the surface $\eta_s$ and is plotted in Fig. 8.

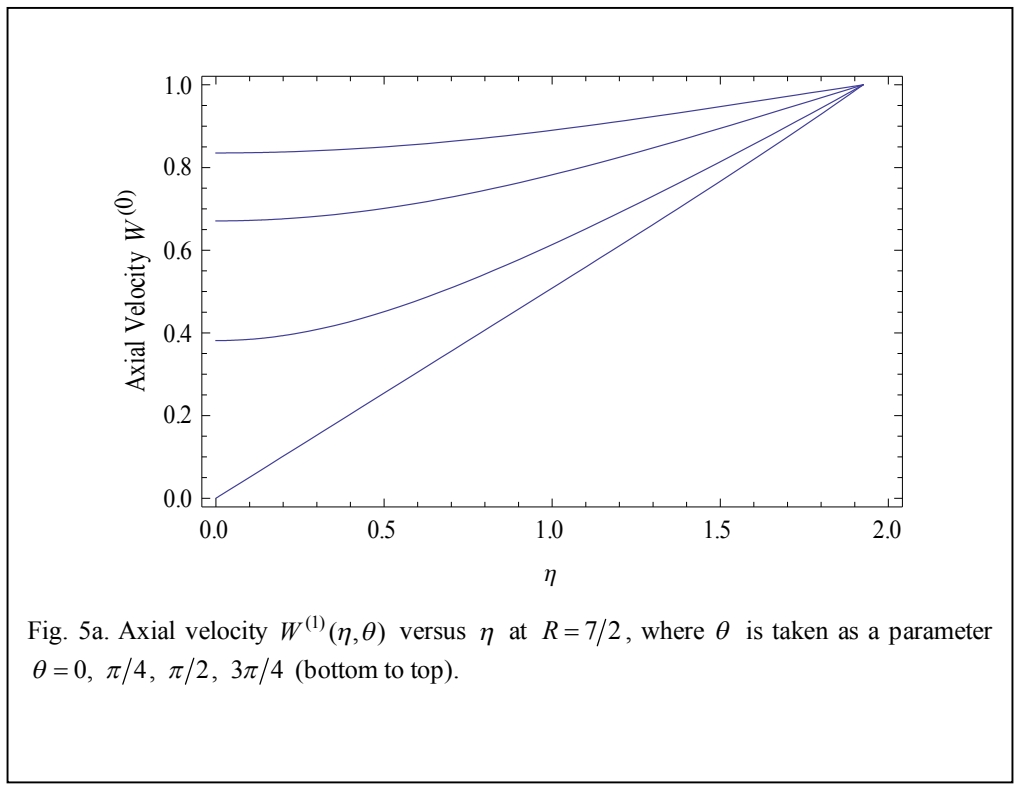

Fig. 5a. Axial velocity $W^{(1)}(\eta,\theta)$ versus $\eta$ at $R=7/2$, where $\theta$ is taken as a parameter $\theta = 0$, $\pi/4$, $\pi/2$, $3\pi/4$ (bottom to top).

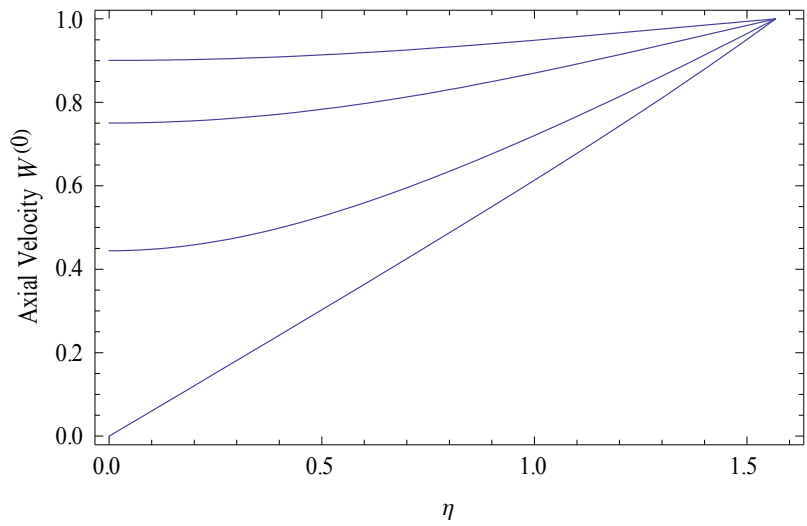

Fig. 5b. Axial velocity $W^{(1)}(\eta,\theta)$ versus $\eta$ at $R = 5/2$, where $\theta$ is taken as a parameter $\theta = 0,\ \pi/4,\ \pi/2,\ 3\pi/4$ (bottom to top).

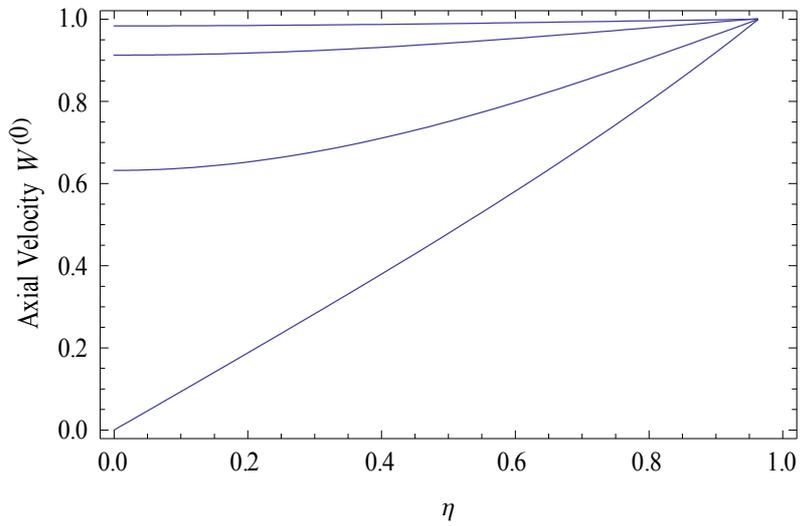

Fig. 5c. Axial velocity $W^{(1)}(\eta,\theta)$ versus $\eta$ at $R = 3/2$, where $\theta$ is taken as a parameter $\theta = 0,\ \pi/4,\ \pi/2,\ 3\pi/4$ (bottom to top).

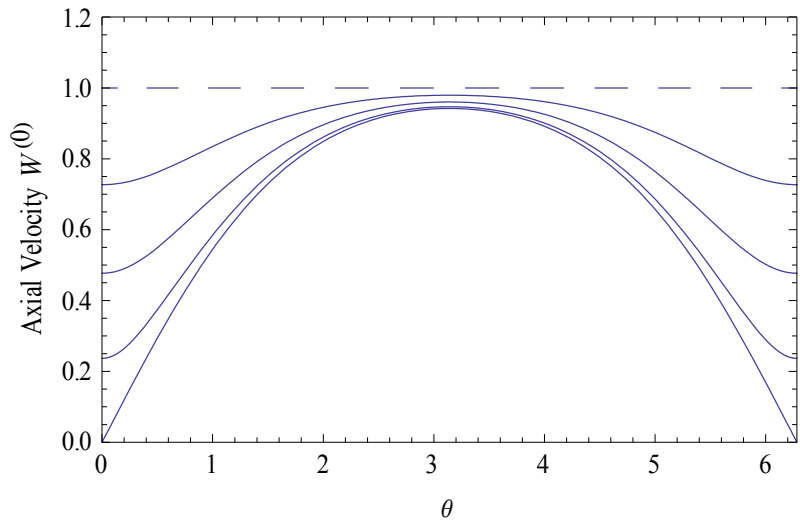

Fig. 6b. Axial velocity $W^{(1)}(\eta,\theta)$ versus $\theta$ at $R = 5/2$, where $\eta$ is taken as a parameter $\eta = \eta_s$, $3\eta_s/4$, $\eta_s/2$, $\eta_s/4$, $0$ (top to bottom).

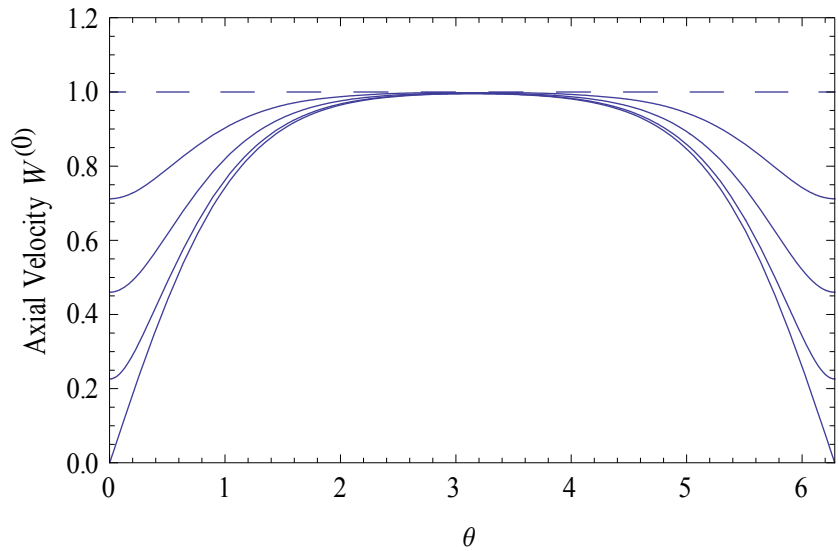

Fig. 6c. Axial velocity $W^{(1)}(\eta,\theta)$ versus $\theta$ at $R = 3/2$, where $\eta$ is taken as a parameter $\eta = \eta_s$, $3\eta_s/4$, $\eta_s/2$, $\eta_s/4$, $0$ (top to bottom).

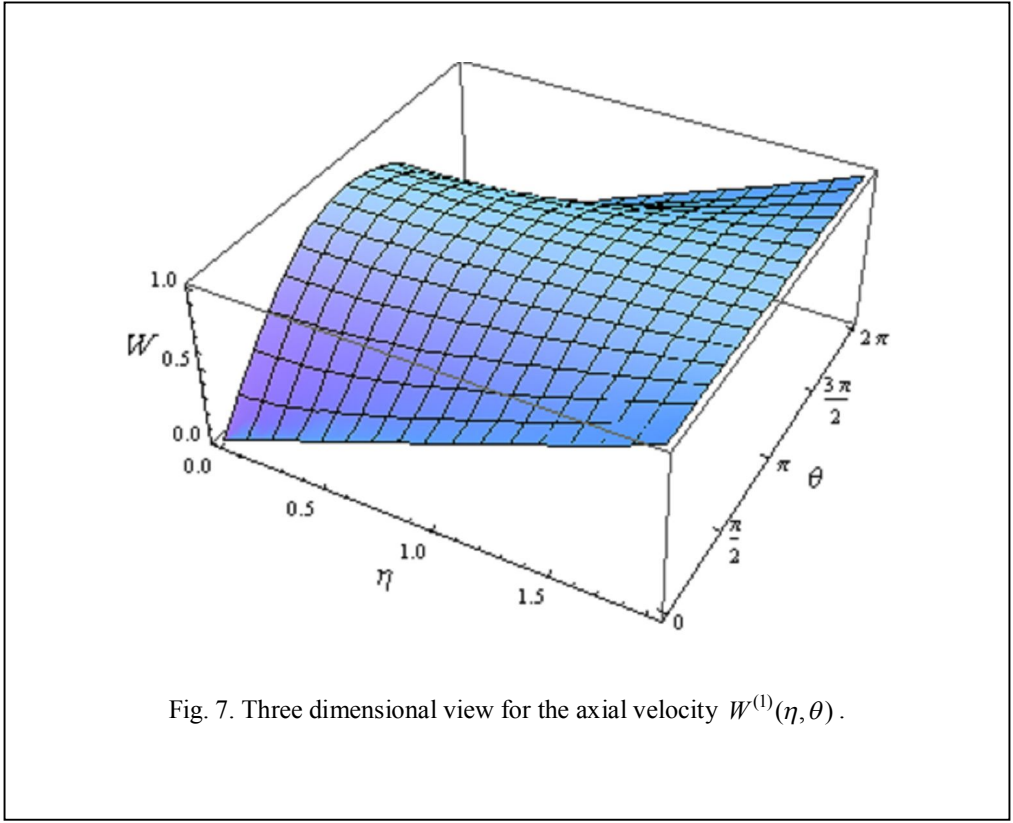

Fig. 7. Three dimensional view for the axial velocity $W^{(1)}(\eta,\theta)$.

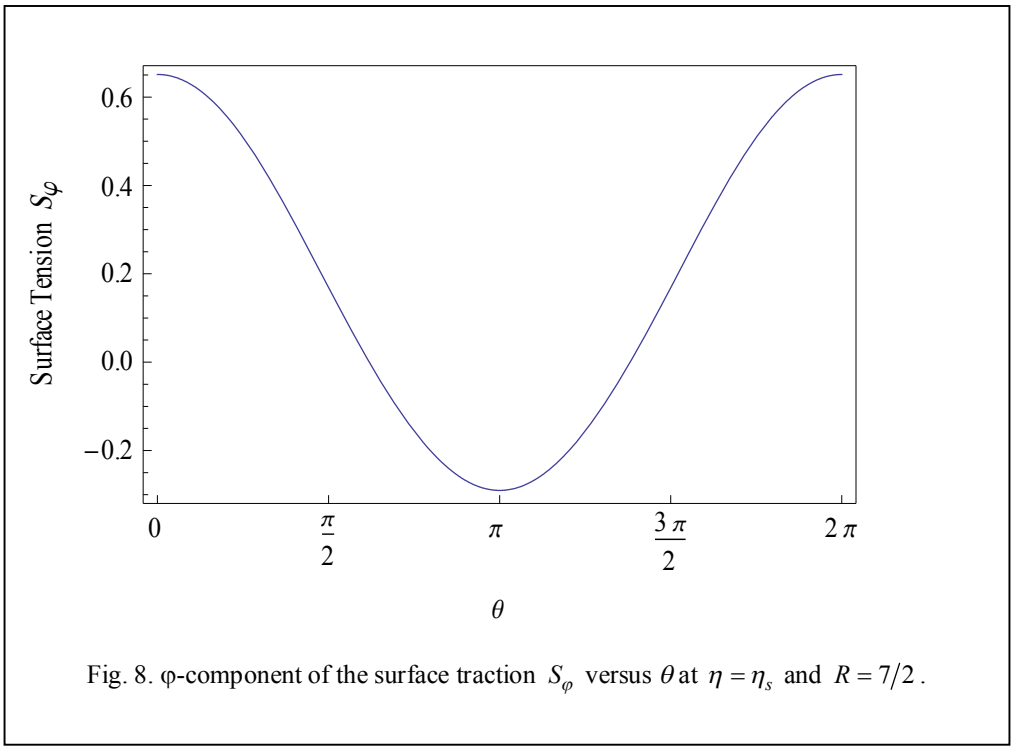

Fig. 8. φ-component of the surface traction $S_\varphi$ versus $\theta$ at $\eta = \eta_s$ and $R = 7/2$.

## 8. Conclusion

The viscoelastic flow around axisymmetric rotation of a rigid torus in an unbounded second order viscoelastic fluid is investigated. The problem is formulated and solved within the frame of slow flow approximation using retarded motion approximation. The equations of motion using the bipolar toroidal coordinate system are formulated. The first order velocity field are determined. The first order velocity component $W^{(1)}(\eta,\theta)$ which lies in the direction of the $\varphi$-coordinate representing the Newtonian flow is obtained but the first order stream function $\psi^{(1)}$ is a vanishing term. The equations of motion of a second order are formulated. The solution of the second order indicates that the axial velocity is vanished while the only nonvanishing term is the second order stream function which will be the subject of a second paper of this series of papers. Laplace's equation of first order velocity $W^{(1)}(\eta,\theta)$ is solved via the usual method of separation of variables. This method shows that, the solution is given in a form of infinite sums over Legendre functions of the first kind. From the obtained solution it is found that, the leading term, $W^{(1)}(\eta,\theta)$ of the velocity. The second-order term shows to be a stream function, $\Psi^{(2)}(\eta,\theta)$, which describes a secondary flow in $\eta\theta$-plane superimposed on the primary flow. The distribution of surface traction which represent the stress vector per unit area at the toroid surface are calculated and discussed.

**REFERENCES**

1. W.R. Dean, "Note on the motion of fluid in a curved pipe," *Phil. Mag.*, vol. 4, pp. 208, 1927.
2. W.R. Dean, "The streamline motion of fluid in a curved pipe," *Phil. Mag.*, vol. 5, pp. 673, 1928.
3. J. Zhang, and B. Zang, "Theoretical and numerical investigation of flow transition in rotating curved annular pipes," *Theoret. Comput. Fluid Dynamics.*, vol. 16, pp. 99, 2002.
4. W.M. Collins and S.C.R. Dennis, "The steady motion of a viscous fluid in a curved tube," *Q.J. Mech. Appl. Math.*, vol. 28, pp. 133, 1975.
5. S.C.R. Dennis, "Calculation of the steady flow through a curved tube using a new finite-difference method," *J. Fluid Mech.*, vol. 99, pp. 449, 1980.
6. M.A. Petrakis, and G.T. Karahalios, "Fluid flow behavior in a curved annular conduit," *Int. J. Non-Linear Mech.*, vol. 34, no. 1, pp. 13, 1999.
7. G. Jayaraman, and K. Tewari, "Flow in catheterized curved artery," *Med. Biol. Eng. Comput.*, vol. 34, no. 5, pp. 720, 1995.
8. R.K. Dash, G. Jayaraman, and K.N. Mehta, "Flow in a catheterized curved artery with stenosis," *J. BioMech.*, vol 32, pp. 49, 1999.
9. S.N. Barua, "On secondary flow in stationary curved pipes," *Q. J. Mech. appl. Math.* vol. 16, pp. 61, 1963.
10. H. Ito, "Laminar flow in curved pipes," *Z. angew. Math. Mech.* vol. 49, pp. 653, 1969.
11. S.C.R. Dennis, and N. Riley, "On the fully developed flow in a curved pipe at high Dean number," *Proc. R. Soc. London*, A 434, pp. 473, 1991.
12. S.A. Berger, L. Talbot, and L.S. Yao, "Flow in curved pipes," *Annu. Rev. Fluid Mech.* vol. 15, pp. 461, 1983.
13. Z.H. Yang and H. Keller, "Multiple laminar flows through curved pipes," *Appl. Numer. Maths.* vol. 2, pp. 257, 1986.

14. J.H. Siggers, and S.L. Waters, "Steady flows in pipes with finite curvature," *Phys. Fluids.* vol. 17, pp. 077102, 2005.
15. J.H. Siggers, and S.L. Waters, "Unsteady flows in pipes with finite curvature," *J. Fluid Mech.* vol. 600, pp. 133, 2008.
16. W.Y. Soh, and S.A. Berger, "Fully developed flow in a curved pipe of arbitrary curvature ratio," *Int. J. Numer. Meth. Fluid.* vol. 7, pp. 733, 1987.
17. W. Jitchote, and A.M. Robertson, "Flow of second-order fluids in curved pipes," *J. Non-Newtonian Fluid Mech.* vol. 90, pp. 91, 2000.
18. V. Coscia, and A. Robertson, "Existence and uniqueness of steady, fully developed flows of second-order fluids in curved pipes," *Math. Models Methods Appl. Sci.* vol. 11, pp. 1055, 2001.
19. H.G. Sharma, and A. Prakash, "Flow of a second-order fluid in a curved pipe," *Indian J. Pure Appl. Math.* vol. 8, pp. 546, 1977.
20. W. Jitchote, and A.M. Robertson, "Flow of second-order fluid in curved pipe," *J. Non-newton. Fluid Mech.* vol. 90, pp. 91, 2000.
21. A.M. Robertson, and S.J. Muller, "Flow of Oldroyd-B fluids in curved pipes of circular and annular cross-section," *Int. J. Non linear Mech.* vol. 31, pp. 3, 1996.
22. Y. Fan, R.I. Tanner, and N. Phan-Thien, "Fully developed viscouss and viscoelastic flows in curved pipes," *J. Fluid Mech.* vol. 440, pp. 327, 2001.
23. C.F. Hsu, and S.V. Patankar, "Analysis of laminar non-Newtonian flow and heat transfer in curved tubes," *AICHE J.* vol. 28, pp. 610, 2004.
24. N. Phan-Thien, and Zheng, "Viscoelastic flow in a curved duct: a similarity solution for the Oldroyd-B fluid," *ZAMP: J. Appl. Math.Phys.* vol. 41, pp. 766, 1990.
25. H.Y. Tsang, and D.F. James, "Reduction of secondary motion in curved tubes by polymer additives," *J. Rheology.* vol. 24, pp. 589, 1980.
26. S. Yanase, N. Goto, and K. Yamamoto, "Dual solutions of the flow through a curved tube," *Fluid Dyn. Res.* vol. 5, pp. 191, 1989.
27. P.M. Morse and H. Feshbach, *Methods of Theoretical Physics*, Part I. New York, McGraw Hill, pp. 997, 1953.
28. P. Moon and D.E. Spencer, *Field Theory for Engineers*, Princeton, NJ. Van Nostrand, 1961.
29. B.D. Coleman, and W. Noll "An approximation theorem for functional with applications in continuum mechanics," *Arch. Ration. Mech. Analysis.* vol. 6, pp. 355, 1960.
30. W.E. Langlois, *Slow viscous flow*, The Macmillan Company, New York, 1964.
31. M. Andrews, "Alternative separation of Laplace's equation in toroidal coordinates and its application to electrostatics," *I. Electrostatics.* vol. 64, pp. 664, 2006.